\documentclass[11pt]{article}

\usepackage{latexsym,amsfonts,amsmath,amssymb,pstricks,wasysym,stmaryrd,enumerate,epsfig}
\usepackage{mathrsfs}
\usepackage{theorem}

\numberwithin{equation}{section}

\newcommand{\be}{\begin{equation}}
\newcommand{\ee}{\end{equation}}
\newcommand{\beu}{\begin{equation*}}
\newcommand{\eeu}{\end{equation*}}
\newcommand{\bea}{\begin{eqnarray}}
\newcommand{\eea}{\end{eqnarray}}
\newcommand{\beaa}{\begin{eqnarray*}}
\newcommand{\eeaa}{\end{eqnarray*}}
\newcommand{\bmx}{\begin{pmatrix}}
\newcommand{\emx}{\end{pmatrix}}

\newcommand{\del}{\partial}

\newcommand{\pd}[1]{{\frac{\del}{\del\smash{ #1}}}}
\newcommand{\cre}{{a^\hc}}

\newcommand{\half}{\frac{1}{2}}

\newcommand{\arctanh}{\tanh^{-1}}

\newcommand{\nn}{\nonumber}

\newcommand{\8}{{\infty}}

\newcommand{\eps}{\epsilon}

\newcommand{\ad}{{\rm ad}}

\newcommand{\hc}{{\dagger}}

\newcommand{\ket}[1]{{\,\left|#1\right>}\,}

\newcommand{\ip}{{\cdot}}
\newcommand{\cas}{{\mathscr C}}

\advance\textheight by \topskip
\begin{document}

\baselineskip 17.5pt
\parindent 18pt
\parskip 8pt

\begin{flushright}
\break

DCPT-07/59

\end{flushright}
\vspace{1cm}
\begin{center}
{\LARGE {\bf Covariant particle statistics and }}

{\LARGE {\bf intertwiners of the $\kappa$-deformed Poincar\'e algebra}}\\[4mm]
\vspace{1.5cm}
{\large  C. A. S. Young\footnote{\texttt{charlesyoung@cantab.net}} and R. Zegers\footnote{\texttt{robin.zegers@durham.ac.uk}}}
\\
\vspace{10mm}

{ \emph{Department of Mathematical Sciences\\ University of Durham\\
South Road, Durham DH1 3LE, UK}}

\end{center}

\vskip 1in
\centerline{\small\bf ABSTRACT}
\centerline{
\parbox[t]{5in}{\small 
To speak about identical particles -- bosons or fermions -- in quantum field theories with $\kappa$-deformed Poincar\'e symmetry, one must have a $\kappa$-covariant notion of particle exchange. This means constructing intertwiners of the relevant representations of $\kappa$-Poincar\'e. We show, in the simple case of spinless particles, that intertwiners exist, and, supported by a  perturbative calculation to third order in $\frac{1}{\kappa}$, make a conjecture about the existence and uniqueness of a certain preferred intertwiner defining particle exchange in $\kappa$-deformed theories.
}}


\vspace{1cm}

\newpage
\section{Introduction and overview}
The so-called $\kappa$-deformation of Minkowski space and the Poincar\'e algebra is an important and much-studied example of a deformation of classical spacetime and its symmetries. The four-dimensional $\kappa$-Poincar\'e algebra was first obtained in \cite{kPoin}, following earlier work on the two-dimensional case \cite{CGST}, by a contraction of a simple quantum group which parallels the usual In\"on\"u-Wigner \cite{IW} contraction $SO(1,4) \rightarrow $ Poincar\'e. 

In terms of the coordinate functions $x^\mu$, $\kappa$-deformation preserves the usual commutativity of the position coordinates but introduces a deformation in the commutators involving time \cite{SZak,DJMTWW}:
\be \left[ x^i, x^j \right] = 0 \,, \qquad \left[ x^0, x^i \right] = \frac{1}{\kappa} x^i \ee
where $\kappa$ is a deformation parameter with dimensions of length. A characteristic effect of $\kappa$-deformation is a modification of the dispersion relation, which becomes\footnote{This formula is expressed in terms of the ``original'' basis of $\kappa$-Poincar\'e, whereas the Casimir (\ref{Caz}) below is in the ``bicrossproduct'' basis \cite{bicross}.}
\be \left(2 \kappa \sinh \frac{E}{2\kappa}\right)^2 - \vec P \ip \vec P\, ;\label{disp}\ee 
This dispersion relation has appeared in other contexts, including  doubly special relativity \cite{DSR,DIKN05,AAC} -- for discussion see \cite{LNdsrVk,Lukierski:2006fv} -- and,
after exchanging $E\leftrightarrow P$, the worldsheet magnons of the AdS/CFT correspondence \cite{AdSCFT}.  
 
The present work is a contribution to the project of formulating quantum field theories with $\kappa$-deformed symmetries. There is an extensive literature on this topic (and on classical field theory): see e.g. \cite{LRZ,AAC,KLM,GW,DFR,Kim:2007ih,AAstar,FS94}. Two broad approaches exist: from one point of view, quantum field theories are constructed, roughly speaking, by quantizing classical field theories; in this approach the non-commutative algebra of coordinates assumes primacy \cite{DJMTWW,Moller,KMLS98,KMS,DMT04}.  Here, on the other hand, we follow the point of view of Weinberg in \cite{Weinberg} that particles, thought of as irreducible representations of the spacetime symmetry algebra, are the fundamental building blocks, with quantum fields and their field equations emerging later.
The first thing one needs, then, is a knowledge of the irreps of $\kappa$-Poincar\'e, i.e. of single-particle states. Fortunately $\kappa$-Poincar\'e and Poincar\'e are known to be isomorphic as algebras (not as Hopf algebras) so they share the same representations \cite{KLMS94,GGM96,KLM01}. Next, to understand  interacting theories, one must be able to take tensor products of representations, i.e. to build multi-particle states. The coalgebra structure of $\kappa$-Poincar\'e (which is genuinely distinct from that of Poincar\'e) allows one to do just this.  
 
In ordinary QFT there is a further key ingredient -- the notion of \emph{identical}, or indistinguishable, particles, with definite exchange statistics. This is most familiarly expressed in terms of the algebra of creation operators which act on the Fock vacuum:
\be \cre(p) \cre(q) \pm \cre(q) \cre(p) = 0 \ee
with $+$ ($-$) for bosons (respectively, fermions). What this really says is that the space of two-particle states is not some tensor product $V\otimes V$ but rather the quotient of this by the operation $\tau$ ($-\tau$) which exchanges the factors:
\be \tau : \ket p \otimes \ket q \mapsto \ket q \otimes \ket p \label{flip}.\ee 
It is important that this definition is covariant, so that the notions of particle exchange and hence identical particles are the same in all frames. This is true since $\tau$ commutes with action of the (co-commutative) Poincar\'e algebra -- that is, it is an intertwiner of representations. 

In this paper we ask: is there an analogue of $\tau$ in the $\kappa$-deformed case? The question is non-trivial because no universal $R$-matrix is known (although c.f. \cite{BKLVY,MSSG}),  so it is necessary to look for intertwiners ``by hand''.\footnote{This is in contrast to another much-studied deformation, $\theta$-deformation, in which the noncommutativity of the coordinates is specified by a constant matrix: $\left[x^\mu,x^\nu\right] = \theta^{\mu\nu}$ \cite{Szabo:2001kg,Douglas:2001ba,strings}. $\theta$-deformation does possess a universal $R$ matrix and indeed is a ``twist'' of the usual commutative case (see e.g. \cite{NgeomGrav}). Because twists are in a certain sense gauge transformations \cite{Majid}, all the relevant structures -- including the algebra of creation operators -- can be obtained in a controlled way, but there has been some debate about whether the resulting theory is physically distinct from the undeformed case \cite{twists}.}  What one \emph{means} by $\kappa$-deformed quantum field theory hinges on whether  suitable intertwiners exist: if there are no such intertwiners then there is no covariant notion of identical particles, and the counting of multi-particle states will differ from the undeformed case; c.f. \cite{AMFock}.
This may indeed be the case for some deformations of Poincar\'e, but it is a radical conceptual departure that, in our view, is physically unattractive and to be avoided if possible; it is better to modify the minimum one can get away with. Our approach is thus closer in spirit to the very interesting papers \cite{DLW1,DLW2}, with the distinction that we shall insist that our notion of particle exchange is covariant, i.e. intertwines the action of the boosts $N_i$ as well as the momentum operators $P_i$. 

In what follows, after briefly recalling the definition of the $\kappa$-Poincar\'e Hopf algebra, we will address the question above, restricting ourselves to the simplest possible case: the two-particle sector of a theory of one species of spinless particle. We find that intertwiners do exist, and also introduce a list of further conditions which we conjecture (based on a perturbative calculation to third order in $\frac{1}{\kappa}$)  specify the \emph{unique} intertwiner naturally associated with $\kappa$-deformed particle-exchange.

\section{Intertwiners for the $\kappa$-Poincar\'e Hopf algebra}

We will work in the bicrossproduct basis, first presented in \cite{bicross}.\footnote{One also often sees the ``modified'' bicrossproduct basis \cite{ALZ,AMFock,DLW1,AAC}, which is defined by $P^{(\text{mod})}_i= e^{\frac{P_0}{2\kappa}} P^{(\text{here})}_i$.}
In this basis the Lorentz algebra is undeformed:
\be \left[ M_i, M_j \right] = \eps_{ijk} M_k, \quad \left[M_i,N_j\right] = \eps_{ijk} N_k, \quad \left[N_i, N_j\right] = -\eps_{ijk} M_k \label{lorentz},\ee
where $M_i$ and $N_i$ generate respectively rotations and boosts. The remaining algebraic relations, involving the generators $P_0$ and $P_i$ of time- and space-translations, are 
\be \left[ P_i , P_0 \right] = 0 ,\quad \left[P_i,P_j\right] = 0\ee 
\be \left[ M_i , P_0 \right] = 0 ,\quad \left[M_i,P_j\right] = \eps_{ijk} P_k \ee
\be \left[ N_i , P_0 \right] = -P_i,\quad \left[N_i,P_j\right] = 
         \delta_{ij} \left(\frac{\kappa}{2}\left(1-e^{-\frac{2P_0}{\kappa}}\right) +\frac{1}{2\kappa} \vec P \ip \vec P \right) 
        - \frac{1}{k} P_i P_j .\ee 
The coalgebra is given by
\be \Delta M_i = M_i \otimes 1 + 1 \otimes M_i, \qquad \Delta N_i = N_i \otimes 1 + e^{-\frac{P_0}{\kappa}} \otimes N_i + \frac{1}{\kappa} \eps_{ijk} P_j \otimes M_k \label{DeltaM}\ee 
\be \Delta P_0 = P_0 \otimes 1 + 1 \otimes P_0, \qquad \Delta P_i = P_i \otimes 1 + e^{-\frac{P_0}{\kappa}} \otimes P_i \label{DeltaP}\ee
and, to complete the Hopf algebra structure, the antipode map is
\be SM_i = -M_i, \qquad SN_i = -e^{\frac{P_0}{k}} N_i + \frac{1}{\kappa} \eps_{ijk} e^{\frac{P_0}{\kappa}} P_j M_k \ee
\be SP_0 = - P_0, \qquad SP_i = -e^{\frac{P_0}{k}} P_i.\ee

These relations define the $\kappa$-Poincar\'e Hopf algebra $\mathcal P_\kappa$. In the limit $\kappa\rightarrow \8$ one recovers the Poincar\'e algebra, and the coproducts collapse to give the usual Leibnitz rule $\Delta J = J\otimes 1 + 1 \otimes J$. 

Single-particle states then, by definition, fall into irreducible representations of $\mathcal P_\kappa$. As usual we can label states by their momentum eigenvalues:
\be P_0 \ket{p_0,\vec p, \sigma} = p_0 \ket{p_0,\vec p,\sigma}, 
 \quad P_i \ket{p_0,\vec p, \sigma} = p_i \ket{p_0, \vec p ,\sigma}.\ee
Here $\sigma$ denotes any further quantum numbers needed to label the states (strictly, part of the definition of \emph{single}-particle states is that $\sigma$ runs over a discrete set of indices only \cite{Weinberg}). In this paper we would like to focus on the simplest case, that of spinless particles; then, since any purely internal degrees of freedom (commuting with $\mathcal P_\kappa$) can be safely ignored, we may drop the index $\sigma$ from now on. 

Let $V$ denote the space of single-particle states, spanned by the $\ket{p_0,\vec p}$. Consider the infinitesimal boost $(1+ \alpha \vec n \ip \vec N)$, where $\alpha\in\mathbb R$ is small and $\vec n\ip \vec n = 1$. We compute its action on $V$ in the usual fashion, dropping terms of $O(\alpha^2)$:
\bea &&  P_i \left(1+ \alpha \vec n \ip \vec N\right) \ket{p_0,\vec p} \\
   &=& \left(1 + \alpha \vec n \ip \vec N\right) \left(1- \alpha \vec n \ip \vec N\right)
      P_i \left( 1+  \alpha \vec n \ip \vec N\right) \ket{p_0,\vec p} \\
&=& \left(1 + \alpha \vec n \ip \vec N\right) 
       \left( P_i + \alpha n_j \left[P_i,N_j\right]\right)  \ket{p_0,\vec p} \\
&=& \left[ p_i - \alpha n_i \left(\frac{\kappa}{2}\left(1-e^{-\frac{2p_0}{\kappa}}\right) +\frac{1}{2\kappa} \vec p \ip \vec p \right) 
        + \frac{\alpha}{\kappa} p_i \vec n \ip \vec p \right] \left(1 + \alpha \vec n \ip \vec N\right) \ket{p_0,\vec p}\eea
and similarly for $P_0$, from which we conclude that 
\be  \left(1+ \alpha \vec n \ip \vec N\right) \ket{p_0,\vec p} 
 = \ket{p_0 - \alpha \vec n \ip \vec p,  p_i - \alpha n_i \left(\frac{\kappa}{2}\left(1-e^{-\frac{2p_0}{\kappa}}\right) +\frac{1}{2\kappa} \vec p \ip \vec p \right) 
        + \frac{\alpha}{\kappa} p_i \vec n \ip \vec p }\ee
up to a possible normalization factor.
In the same way, we may compute the action of infinitesimal boosts on two-particle states in $V\otimes V$:
\bea &&(1\otimes P_i) \left(1\otimes 1+ \alpha \vec n \ip \Delta \vec N\right) 
                      \ket{r_0,\vec r}\otimes \ket{s_0, \vec s} \\
&=&  \left(1\otimes 1+ \alpha \vec n \ip \Delta \vec N\right) \left( 1\otimes P_i + \alpha n_j \left[1\otimes P_i, \Delta N_j\right] \right)  \ket{r_0,\vec r}\otimes \ket{s_0, \vec s} \\
&=&  \left[ s_i - \alpha n_i e^{-\frac{r_0}{\kappa}} \left( \frac{\kappa}{2} \left(1-e^{-\frac{2s_0}{\kappa}}\right) + \frac{1}{2\kappa} \vec s \ip \vec s \right) + \frac{\alpha}{\kappa} e^{-\frac{r_0}{\kappa}} s_i \vec n\ip \vec s + \frac{\alpha}{\kappa}r_i \vec n \ip \vec s - \frac{\alpha}{\kappa} n_i \vec r \ip \vec s \right]\nn\\ &&\qquad\times \left(1\otimes 1+ \alpha \vec n \ip\Delta \vec N\right) \ket{r_0,\vec r}\otimes \ket{s_0, \vec s},\eea
and similarly for $P_i \otimes 1$, $P_0\otimes 1$ and $1\otimes P_0$, yielding
\bea &&\left(1\otimes 1+ \alpha \vec n \ip \Delta \vec N\right) 
                      \ket{r_0,\vec r}\otimes \ket{s_0, \vec s}\label{2p}\\
=&& \ket{r_0 - \alpha \vec n \ip \vec r,  r_i - \alpha n_i \left(\frac{\kappa}{2}\left(1-e^{-\frac{2r_0}{\kappa}}\right) +\frac{1}{2\kappa} \vec r \ip \vec r \right) 
        + \frac{\alpha}{\kappa} r_i \vec n \ip \vec r } \nn\\
&\otimes& \ket{s_0 - \alpha e^{-\frac{r_0}{\kappa}} \vec n \ip \vec s,
                s_i - \alpha n_i e^{-\frac{r_0}{\kappa}} \left( \frac{\kappa}{2} \left(1-e^{-\frac{2s_0}{\kappa}}\right) + \frac{1}{2\kappa} \vec s \ip \vec s \right) + \frac{\alpha}{\kappa} e^{-\frac{r_0}{\kappa}} s_i \vec n\ip \vec s + \frac{\alpha}{\kappa}r_i \vec n \ip \vec s - \frac{\alpha}{\kappa} n_i \vec r \ip \vec s}\nn \eea
We choose to set the normalization factor to unity here.

As an aside, let us note that the antipode map is not used here, which is natural given that there are no conjugate representations in sight. It is in fact true that the Hopf-algebraic notion of the adjoint action of $N_i$ on $P_j$ agrees with the usual un-deformed one:
\be \ad_{N_i} P_j := SN_{j(1)} P_i N_{j(2)} = \left[ P_i, N_j \right] \label{qic}.\ee
Consequently, and as one might expect, the \emph{total} momentum $p_\mu$ of an $m$-particle state transforms, under the action of a boost on the state, in the same fashion as does the vector $P_\mu$ of algebra generators under the adjoint action of the boost.\footnote{To see this explicitly, note that $\ad_{\alpha \vec n \ip \vec N} P_\mu$ evaluates to some power series in the generators $P_\nu$. Call this series $A_\mu(P)$. $\Delta^m A(P) = A(\Delta^m P)$, since $\Delta$ is an algebra homomorphism. Now on a state $\ket{} \in V^{\otimes m}$ such that $\Delta^m P_\mu \ket{} = p_\mu\ket{}$ one has
\be (\Delta^m P_\mu) (1^{\otimes m} + \alpha \vec n \ip \Delta^m   \vec N) \ket{} 
  = (1^{\otimes m} + \alpha \vec n \ip \Delta^m \vec N) \left( \Delta^m P_\mu + \left[ \Delta^m P_\mu ,\Delta^m \alpha \vec n \ip  \vec N\right] \right) \ket{}\ee
so the change in total momentum is given by
\bea \delta p_\mu\ket{} &:=& \left[ \Delta^m P_\mu , \Delta^m \alpha \vec n \ip \vec N \right]\ket{}  =\Delta^m [P_\mu, \alpha \vec n \ip \vec N]\ket{} \\&=& \Delta^m \ad_{\alpha \vec n \ip\vec N} P_\mu\ket{} = \Delta^m A_\mu(P)\ket{} = A_\mu(\Delta^m P) \ket{} = A_\mu(p)\ket{}\eea
which establishes that $\delta p_\mu = A_\mu(p)$, i.e. that indeed $p_\mu$ transforms like $P_\mu$.}
But (\ref{qic}) holds only by virtue of the specific form of the coproduct and antipode in $\mathcal P_\kappa$; it is certainly not generic, and in particular $\ad_{\Delta N_j} \Delta P_i \neq \left[\Delta P_i, \Delta N_j\right]$. Thus it is important to remember that the total momentum of a boosted two-particle state is \emph{not} equal to
\be \ad_{1\otimes 1 + \Delta \alpha \vec n \ip \vec N} \Delta P_\mu \ee
evaluated on the original state.

Our goal now is to find maps
\be \tau: V \otimes V \rightarrow V \otimes V \ee
that intertwine (i.e. commute with) the action of $\mathcal P_\kappa$:
\be \tau\,\,\Delta J = \Delta J\,\, \tau \quad \text{for all} \quad J\in \mathcal P_\kappa,\ee
and that reduce in the limit $\kappa\rightarrow \8$ to exchange of factors in the tensor product:
\be \tau_{\kappa=\8} \,: \ket r \otimes \ket s \mapsto \ket s \otimes \ket r \label{flip}\ee
(which is of course an intertwiner of classical Poincar\'e, because the coproducts are co-commutative). 
As in \cite{DLW1}, we will seek maps of the form 
\be \tau : \ket{r_0, \vec r} \otimes \ket{s_0,\vec s} \rightarrow 
 \ket{\smash{f_0(r,s),\vec f(r,s)}} \otimes \ket{ g_0(r,s), \vec g(r,s)}.\label{taudef}\ee
This is clearly not the most general form conceivable -- we could take a linear superposition of basis states on the right hand side -- but it will turn out to work.   
First let us demand that $\tau$ commute with boosts. On the one hand we have, following the rule of (\ref{2p}),
\bea &&\left(1\otimes 1+ \alpha \vec n \ip \Delta \vec N\right) \tau  \ket{r_0, \vec r} \otimes \ket{s_0,\vec s} \\
=&& \ket{f_0 - \alpha \vec n \ip \vec f,  f_i - \alpha n_i \left(\frac{\kappa}{2}\left(1-e^{-\frac{2f_0}{\kappa}}\right) +\frac{1}{2\kappa} \vec f \ip \vec f \right) 
        + \frac{\alpha}{\kappa} f_i \vec n \ip \vec f } \label{Nt}\\
&\otimes& \ket{g_0 - \alpha e^{-\frac{f_0}{\kappa}} \vec n \ip \vec g,
                g_i - \alpha n_i e^{-\frac{f_0}{\kappa}} \left( \frac{\kappa}{2} \left(1-e^{-\frac{2g_0}{\kappa}}\right) + \frac{1}{2\kappa} \vec g \ip \vec g \right) + \frac{\alpha}{\kappa} e^{-\frac{f_0}{\kappa}} g_i \vec n\ip \vec g + \frac{\alpha}{\kappa}f_i \vec n \ip \vec g - \frac{\alpha}{\kappa} n_i \vec f \ip \vec g}\nn \eea
On the other hand, Taylor-expanding $f$ and $g$ to first order in $\alpha$, we have that
\bea &&\tau \left(1\otimes 1+ \alpha \vec n \ip \Delta \vec N\right)  \ket{r_0, \vec r} \otimes \ket{s_0,\vec s} \\
&=& \ket{\left(1+\alpha \vec n \ip \vec D\right)f_0, \left(1+\alpha \vec n \ip \vec D\right) f_j}
  \otimes \ket{\left(1+\alpha \vec n \ip \vec D\right) g_0, \left(1+\alpha \vec n \ip \vec D\right) g_j}\label{tN}\eea
where the differential operator $D_i$ is defined to be
\bea D_i &:=& -r_i \pd{r_0} - \left(\frac{\kappa}{2}\left(1-e^{-\frac{2r_0}{\kappa}}\right) +\frac{1}{2\kappa} \vec r \ip \vec r \right) \pd{r_i} + \frac{1}{\kappa} r_i r_j \pd{r_j}  \nn\\
&& - e^{-\frac{r_0}{\kappa}} s_i \pd{s_0} - e^{-\frac{r_0}{\kappa}} \left( \frac{\kappa}{2} \left(1-e^{-\frac{2s_0}{\kappa}}\right) + \frac{1}{2\kappa} \vec s \ip \vec s \right)\pd{s_i} + \frac{1}{\kappa} e^{-\frac{r_0}{\kappa}} s_i s_j \pd{s_j} \nn\\ && + \frac{1}{\kappa} r_j s_i \pd{s_j} - \frac{1}{\kappa} \vec r \ip \vec s \pd{s_i}.\label{D}\eea 

Comparing (\ref{Nt}) and (\ref{tN}) one sees that $\tau$ commutes with the action of boosts if and only if the following set of PDEs are satisfied: 
\bea D_i f_0 &=& -f_i \label{PDE1}\\
     D_i f_j &=& -\delta_{ij} \left(\frac{\kappa}{2}\left(1-e^{-\frac{2f_0}{\kappa}}\right) +\frac{1}{2\kappa} \vec f \ip \vec f \right) 
        + \frac{1}{\kappa} f_i f_j \label{PDE2}\\
     D_i g_0 &=& -g_i e^{-\frac{f_0}{\kappa}} \label{PDE3}\\
     D_i g_j &=& -\delta_{ij} e^{-\frac{f_0}{\kappa}} \left( \frac{\kappa}{2} \left(1-e^{-\frac{2g_0}{\kappa}}\right) + \frac{1}{2\kappa} \vec g \ip \vec g \right) + \frac{1}{\kappa} e^{-\frac{f_0}{\kappa}} g_i g_j + \frac{1}{\kappa}f_j g_i - \frac{1}{\kappa} \delta_{ij} \vec f \ip \vec g\label{PDE4}.\eea
Of course we also require that $\tau$ commute with spatial rotations. This should follow automatically from the equations above because every rotation generator can be written as a bracket of boosts. It is a useful check to verify that this is so. First, one can compute 
\be \left[ D_i, D_j\right] = r_i \pd{r_j} - r_j \pd{r_i}  + s_i \pd{s_j} - s_j \pd{s_i}\ee
and recognize the right hand side as the correct realization of the rotation generator $-\eps_{ijk} M_k$ on functions of two momenta, given that the algebra and coalgebra of the $M_i$ is entirely classical. Second, it follows from (\ref{PDE1}-\ref{PDE4}) that
\be \left[ D_i, D_j\right] f_0 = 0,\qquad\left[ D_i, D_j\right] f_k = \delta_{jk} f_i - \delta_{ik} f_j \ee
\be \left[ D_i, D_j\right] g_0 = 0,\qquad\left[ D_i, D_j\right] g_k = \delta_{jk} g_i - \delta_{ik} g_j \ee
so that indeed $f_0$ and $g_0$ are singlets of $so(3)$  while $\vec f$ and $\vec g$ are vectors. 

The final requirement for $\tau$ to be an intertwiner of $\mathcal P_\kappa$ is that it should commute with the translation generators $P_\mu$. In view of the coproduct (\ref{DeltaP}), this is the case if and only if 
\bea f_0 + g_0 &=& r_0 + s_0\label{iP0}\\
    f_i + e^{-\frac{f_0}{\kappa}} g_i &=& r_i + e^{-\frac{r_0}{\kappa}} s_i\, , \label{iPi}\eea
as previously noted in \cite{DLW1}.
The crucial point now is that (\ref{PDE1}-\ref{PDE4}) and (\ref{iP0}-\ref{iPi}) \emph{are} a consistent set of equations. We have verified that, when $g$ is eliminated in favour of $f$ using  (\ref{iP0}-\ref{iPi}), (\ref{PDE3}-\ref{PDE4}) are equivalent to (\ref{PDE1}-\ref{PDE2}). 

The problem of finding intertwiners therefore reduces to finding solutions
\be f_0(r_0,\vec r,s_0, \vec s), \qquad  \vec f(r_0,\vec r,s_0, \vec s) \ee
of the partial differential equations (\ref{PDE1}-\ref{PDE2}). In the following, we will adopt a hands-on approach, and look for solutions order-by-order in $\frac{1}{\kappa}$, which will lend insight into what extra conditions must be given to specify a unique suitable solution. 

\section{Solution: Perturbative Approach}\label{pert}
The operators $D_i$ have well-defined expansions in inverse powers of $\kappa$; we will assume that $f(r,s)$ and $g(r,s)$ do too. The need to recover the correct classical intertwiner (\ref{flip}) in the $\kappa\rightarrow \8$ limit fixes the leading-order behaviour, and we write
\be f_0 = s_0 + \frac{1}{\kappa} f^{(1)}_0+\frac{1}{\kappa^2} f^{(2)}_0 + \dots,\qquad f_i = s_i + \frac{1}{\kappa} f^{(1)}_i +\frac{1}{\kappa^2} f^{(2)}_i + \dots .\ee
At first order in $\frac{1}{\kappa}$ the equations (\ref{PDE1}-\ref{PDE2}) are then
\bea D^{(0)}_i f^{(1)}_0 &=& -f^{(1)}_j + r_0 s_j\\
     D^{(0)}_i f^{(1)}_j &=& -\delta_{ij} f^{(1)}_0 - \delta_{ij} \left( r_0 s_0 - \vec r \ip \vec s\right) + r_j s_i \eea
where
\be D^{(0)}_i = -r_i \pd{r_0} -r_0 \pd{r_i}  -s_i \pd{s_0} -s_0 \pd{s_i} \ee
is the leading term in $D_i$. $f^{(1)}$ must be quadratic in $r,s$ and transform correctly under rotations. This leaves only a few possibilities and, by inspection, one finds that there is a unique solution, namely
\be f^{(1)}_0 = \vec r \ip \vec s ,\qquad f^{(1)}_i = r_i s_0 .\ee
It is then possible to continue to second order in $\frac{1}{\kappa}$, and in fact order-by-order as far as one wishes: at any given order $\kappa^{-m}$ one will find a set of PDEs linear in $f^{(m)}$, and with source terms, all of degree $m+1$ in the momenta, arising from the (by then known) action of $D-D^{(0)}$ on $f$ up to order $\kappa^{-m}$.  
However, there is generally some freedom in this procedure, and this is first seen at order $\frac{1}{\kappa^2}$, where we have found that the general solution is
\bea f^{(2)}_0 &=& Ar_0(\vec r \ip \vec r - r_0 r_0) + Bs_0(\vec s \ip \vec s- s_0s_0)\label{f20}\\
&&\nn\quad{} + (1-C_1-C_2) r_0\vec r \ip \vec s + C_1s_0 \vec r \ip \vec r + C_2 s_0 r_0 r_0 \\
&&\nn\quad{}- (D_1 + D_2) s_0 \vec r \ip \vec s+ D_1r_0\vec s \ip \vec s +D_2r_0s_0s_0 ,\\
     f^{(2)}_i &=&\label{f2i}
 A r_i( \vec r \ip \vec r-r_0r_0)+ Bs_i (\vec s \ip \vec s - s_0s_0) \\&&\quad{} +(1-C_1-C_2)r_i \vec r \ip \vec s + (C_1 + C_2) r_i r_0 s_0 + (C_1-\half) s_i (\vec r \ip \vec r-r_0 r_0)\nn\\&&\quad{}- (1+D_1+D_2) s_i \vec r \ip \vec s + (D_1 + D_2) s_i r_0 s_0 + (D_1 +\half) r_i \vec s \ip \vec s - (1+D_1) r_i s_0 s_0  \nn\eea
for arbitrary coefficients $A,B,C_1,C_2,D_1,D_2$.

We want to arrive at a \emph{unique} intertwiner $\tau$ defining the notion of exchange of identical particles in $\kappa$-deformed theories, so we should use some other input to eliminate this freedom. One obvious thing we can do, in the spirit of keeping as much of the usual structure as possible, is to demand that $\tau^2=1$. Given the definition of $\tau$ in (\ref{taudef}), this means
\be f\left(f(r,s), g(r,s)\right) = r, \qquad g\left( f(r,s), g(r,s) \right) = s,\ee
which, combined with (\ref{iP0}-\ref{iPi}), yields a new condition on $f$. Fortunately this condition is satisfied to order $\frac{1}{\kappa}$, where we have no freedom. At order $\frac{1}{\kappa^2}$ it turns out to produce
\bea f^{(2)}_0(r,s) - f^{(2)}_0(s,r) &=& s_0 \vec r \ip \vec r - r_0 \vec s \ip \vec s \\
     f^{(2)}_i(r,s) - f^{(2)}_i(s,r) &=& \half r_0 r_0 s_i - \half s_0 s_0 r_i 
                                       + \vec r \ip \vec s r_i - \vec r \ip \vec s s_i
                                       + r_0 s_0 r_i - r_0 s_0 s_i, \eea
which reduces the number of free parameters in the solution above from 6 to 3 by forcing
\be A=B,\quad C_1=1+D_1, \quad C_2=D_2 .\label{fromtau2}\ee 

There is a further important constraint on the form of $\tau$.  As in the usual undeformed case, the full space $V$ of spin zero single-particle states, spanned by the modes $\ket{p_0,\vec p}$, falls into a disjoint union of irreps of $\mathcal P_\kappa$, labelled by their values of the mass Casimir
\be \cas = \left(2\kappa \sinh \frac{P_0}{2\kappa} \right)^2 - e^{\frac{P_0}{\kappa}} \vec P \ip \vec P \quad\underset{\kappa\rightarrow\8}\longrightarrow\quad P_0 P_0 - \vec P \ip \vec P=P_\mu P^\mu.\label{Caz}\ee
Thus far we have not been careful about labelling the irrep to which a given ket belongs, but it now becomes important. 
States $\ket r \otimes \ket s$ in the tensor product have three invariant labels, namely
\be  \cas \otimes 1,\quad 1 \otimes \cas  ,\quad \Delta \cas  \label{2labels}\ee
(giving, in the $\kappa\rightarrow \8$ limit, the  two masses $r^\mu r_\mu$ and $s^\mu s_\mu$ and the impact parameter $\left(r+s\right)^\mu \left(r+s\right)_\mu$). On physical grounds, it is reasonable to assume that \emph{identical} particles, in the sense we are seeking to define, must have equal masses, and that these masses are not altered by the operation of exchanging them. That is to say, if we write $\cas(p)$ for the value of the mass Casimir evaluated on $\ket p$, i.e. 
\be\cas(p) \ket p = \cas \ket p, \ee 
then in (\ref{taudef}) we should assume (or, in the case of $f$ and $g$, demand) that
\be \cas (r) = \cas (s) = \cas (f) =\cas (g) \label{masses} .\ee
(Note that of course $\tau$ commutes with $\Delta \cas$ by construction and so automatically preserves the third label in (\ref{2labels}) above.)

Expanding order by order one finds that this condition (\ref{masses}) too is already satisfied to order $\frac{1}{\kappa}$  and produces a new constraint at order $\frac{1}{\kappa^2}$:
\be f^{(2)}_0(r,s) s_0 - \vec f^{(2)}(r,s) \ip \vec s + \half \left(\vec r \ip \vec s\right)^2 -\half\vec r \ip \vec r s_0 s_0- \half (\vec r \ip \vec s) (\vec s \ip \vec s) - \vec r \ip \vec s s_0 s_0 = 0.\ee 
Somewhat remarkably, at least on the face of it, this constraint is compatible with the solution (\ref{f20}-\ref{f2i}) and relations (\ref{fromtau2}). Equating the coefficients of the various terms in $r,s$ gives a large but highly redundant collection of equations, to which there is a single solution for the remaining unknowns:
\be A=0, \quad C_1=\half, \quad C_2=0.\ee
Further, it may be verified that at this order the function $g(r,s)$, determined by (\ref{iP0}-\ref{iPi}), also has the same value of the Casimir, i.e. that the final equality in (\ref{masses}) holds.

We have repeated the steps above at the next order, $\frac{1}{\kappa^3}$, using \textsc{Maple} to help with the rather lengthy calculations. It turns out, even more strikingly than at second order, that the naively overdetermined sets of equations which arise are in fact consistent and have exactly one solution. This solution is
\bea f_0(r,s) &=& s_0 + \frac{1}{\kappa} \vec r \ip \vec s 
        +\frac{1}{2\kappa^2}\left(s_0 \vec r \ip \vec r - r_0 \vec s \ip \vec s + r_0 \vec r \ip \vec s + s_0 \vec r \ip \vec s\right)  \\
   \nn &&{}+ \frac{1}{4\kappa^3}\left( \vec r \ip \vec s \, \vec r \ip \vec r + \vec r \ip \vec s \,\vec s \ip \vec s + 2 r_0 s_0 \vec r \ip \vec r - 2 r_0 s_0 \vec r \ip \vec s -2 r_0 s_0 \vec s \ip \vec s  \right) + \dots\\ &&\nn\\
     f_i(r,s) &=& s_i + \frac{1}{\kappa} r_i s_0
        +\frac{1}{2\kappa^2} \left( r_i \vec r \ip \vec s - s_i \vec r \ip \vec s + r_i r_0 s_0 - s_i r_0 s_0 - r_i s_0 s_0 \right) \\
 &&{}+ \frac{1}{4\kappa^3} \Big( r_i \left( 2 r_0 \vec r \ip \vec s - r_0 \vec s \ip \vec s - 2 r_0 s_0 s_0 + s_0 \vec r \ip \vec r - 2 s_0 \vec r \ip \vec s + \textstyle{\frac{2}{3}}s_0 s_0 s_0 \right) \nn\\\nn && \quad\qquad{} + s_i \left( -2 r_0 \vec r \ip \vec s + r_0 \vec s \ip \vec s - s_0 \vec r \ip \vec r\right) \Big) + \dots \eea \bea
     g_0(r,s) &=& r_0 - \frac{1}{\kappa} \vec r \ip \vec s
       + \frac{1}{2\kappa^2} \left( r_0 \vec s \ip \vec s - s_0 \vec r \ip \vec r - r_0 \vec r \ip \vec s - s_0 \vec r \ip \vec s \right) \\
   && {}- \frac{1}{4\kappa^3}\left( \vec r \ip \vec s \, \vec r \ip \vec r + \vec r \ip \vec s \, \vec s \ip \vec s +2 r_0 s_0 \vec r \ip \vec r - 2 r_0 s_0 \vec r \ip \vec s -2 r_0 s_0 \vec s \ip \vec s \right) +\dots\nn\\&&\nn\\
     g_i(r,s) &=& r_i - \frac{1}{\kappa} s_i r_0
       + \frac{1}{2\kappa^2} \left( r_i \vec r \ip \vec s + s_i \vec r \ip \vec s - r_i r_0 s_0 - s_i r_0 s_0 + s_i r_0 r_0 \right) \\
&&{}+\frac{1}{4\kappa^3} \Big( s_i \left( 2s_0 \vec r \ip \vec s + s_0 \vec r \ip \vec r +2 s_0 r_0 r_0  -  r_0 \vec s \ip \vec s -2  r_0 \vec r \ip \vec s - \textstyle{\frac{2}{3}} r_0 r_0 r_0 \right) \nn\\
&&\nn\quad \qquad{}+ r_i \left( 2 s_0 \vec r \ip \vec s  + s_0 \vec r \ip \vec r - r_0 \vec s \ip \vec s \right) \Big) + \dots \, . \eea

To summarize, to order $\frac{1}{\kappa^3}$ these are the unique functions $f$ and $g$ such that the map $\tau$ in (\ref{taudef}) has the following properties:
\begin{enumerate}
\item $\tau$ is covariant, i.e. commutes with the action of $\mathcal P_\kappa$
\item $\tau$ reduces to the usual map flipping the factors in the $\kappa\rightarrow \8$ limit 
\item $\tau$ squares to the identity \label{t21}
\item $\tau$ preserves the masses of particles.
\end{enumerate}
We will therefore take this to be the complete list of defining properties of the particle-exchange map, and make the following
 
\begin{center}\textbf{Conjecture}: \emph{Such a map $\tau$ exists and is unique, to all orders in $\frac{1}{\kappa}$.}\end{center}

\noindent We have no proof of this, but if it were false then the fact that it works in such an intricate way at the first few orders would be a strange accident; in our view it is much more plausible that the conjecture holds, to all orders, by virtue of some deeper argument that remains to be found.

\section{Solution: Exact Approach}\label{exact}
In light of the discussion above, it is clearly important to solve the equations (\ref{PDE1}-\ref{PDE2}) exactly, to all orders in $\frac{1}{\kappa}$. 
In this section we make a certain amount of progress in this direction, at least in the simpler two-dimensional case, but unfortunately we will not be able to write solutions in a form which gives us sufficient control over the condition (\ref{t21}) in the list above.

It follows from the condition $\cas(f) = \cas(r)$ in (\ref{masses}) that
\be -\left( \frac{\kappa}{2} \left( 1- e^{-\frac{2f_0}{\kappa}}\right) + \half \vec f \ip \vec f \right)
= \frac{1}{2\kappa} \cas(r) e^{-\frac{f_0}{\kappa}} - \kappa + \kappa   e^{-\frac{f_0}{\kappa}}\ee
which considerably simplifies the right-hand side of (\ref{PDE2}). There then exists a choice of new variables
\bea \phi_0 &=&  e^{\frac{f_0}{\kappa}} -1- \frac{\cas(r)}{2\kappa^2}  \\
    \phi_i &=& \frac{1}{\kappa} e^{\frac{f_0}{\kappa}} f_i ,\eea
in which (\ref{PDE1}-\ref{PDE2}) are linear and homogeneous:
\bea D_i \phi_0 &=& -\phi_i \label{lpde1}\\ 
    D_i \phi_j &=& - \delta_{ij}\phi_0.\label{lpde2}\eea
This type of transformation has been seen before: it was shown in \cite{KLMS94,GGM96} that, by  similar changes of variables, all trace of the $\kappa$-deformation can be removed from the algebraic relations $\kappa$-Poincar\'e. (The deformation persists, of course, and can still be seen in the coalgebra, which becomes very unwieldy.) And indeed (\ref{lpde2}) says nothing but that, under the action of the Lorentz algebra on its parameters $r$ and $s$, $\phi_\mu(r,s)$  transforms as a classical Lorentz four-vector: 
\be \phi_\mu(  g \triangleright (r,s) ) = \Lambda(g)_\mu{}^\nu \phi_\nu(r,s) \ee
where $\Lambda$ is the fundamental matrix representation.

We can make, similarly, a new choice of the independent variables, $(r,s)\mapsto (\rho,\sigma)$
\bea
\rho_0 &=&  e^{\frac{r_0}{\kappa}} -1- \frac{\cas(r)}{2\kappa^2}  \qquad\qquad \rho_i = \frac{1}{\kappa} e^{\frac{r_0}{\kappa}} r_i , \\
\sigma_0 &=& e^{\frac{s_0}{\kappa}} -1- \frac{\cas(s)}{2\kappa^2} \qquad\qquad \sigma_i = \frac{1}{\kappa} e^{\frac{s_0}{\kappa}} s_i ,
\eea
in terms of which the operators $D_i$, (\ref{D}), read
\be
D_i = - \rho_i \pd{\rho_0} - \rho_0 \pd{\rho_i} + \frac{1}{\rho_0 + 1+\frac{\cas(r)}{2\kappa^2}} \left(-\sigma_i \pd{\sigma_0} -\sigma_0 \pd{\sigma_i} + \rho_j \sigma_i \pd{\sigma_j} - \vec \rho \ip \vec \sigma \pd{\sigma_i}  \right) .\label{D2}
\ee
Were it not for the last two terms in the bracket and the overall coefficient of the latter, this expression would coincide exactly with the usual differential form of the undeformed two-particle boost.

An explicit solution of (\ref{lpde1}-\ref{lpde2}) can be obtained in the two dimensional case.
In two dimensions there is only one boost operator, $D$, and its expression is given in (\ref{D2}) with $i=j=1$. The last two terms in the bracket cancel and we are left with a simpler expression which, up to the overall coefficient of the bracket, looks a lot like the undeformed two-particle boost operator. A more convenient parametrisation is then obtained by choosing a set of momentum space coordinates adapted to the mass shell:
\bea
\rho_0 &=& m \cosh \theta, \qquad \rho_1 = m \sinh \theta \\
\sigma_0 &=& m \cosh \psi, \qquad \sigma_1 = m \sinh \psi .
\eea
We have chosen the same ``mass'' $m$ for both sets of variables as we are considering identical particles. The value of $m$ is related to the common value of the Casimir by
\be
m^2 = \frac{\cas(r)}{\kappa^2} \left (1 + \frac{\cas(r)}{4\kappa^2} \right )  
\ee
(this is most easily seen in the rest frame of, say, $\rho$). In terms of the rapidities, the differential operator $D$ reads
\be
D = -\pd{\theta}  - \frac{1}{m\cosh \theta +1 + \frac{\cas(r)}{2\kappa^2}} \pd{\psi} ,
\ee
Equations (\ref{lpde1}-\ref{lpde2}) thus reduce to
\be
\left [ \left ( m \cosh \theta +1 +\frac{\cas(r)}{2\kappa^2}\right ) \left (\pd{\theta} \mp 1 \right ) + \pd{\psi} \right ] \phi_\pm =0 ,
\ee
where we have set $\phi_\pm = \phi_0 \pm \phi_1 $. These equations admit a set of solutions in closed form:
\be
\phi_\pm = e^{\pm \theta} F_\pm \left [\psi + 2 \arctanh \left ( \left (m-1-\frac{\cas(r)}{2\kappa^2} \right )  \tanh \frac{\theta}{2} \right )  \right ] ,
\ee
parametrised by the arbitrary functions of one variable $F_\pm$. We have thus obtained the intertwiners $\tau$ in two dimensions exactly. However, applying the extra condition $\tau^2=1$ results in implicit equations for $F_\pm$ that we have not been able to solve. 

What appears to be needed is a reformulation of the defininition (\ref{taudef}) of the intertwiner $\tau$ which includes the condition $\tau^2=1$ in a natural way from the outset. We expect that this should be possible by choosing a new parameterization of the space of two-particle states in which the action of $\tau$ is simple, and we hope to return to this question in a future work.

\vspace{1cm}
\emph{Acknowledgements} C.Y. is grateful to the Leverhulme trust for financial support. R.Z. is supported by an EPSRC postdoctoral fellowship.

\end{document}